\documentclass{pic2012}
\def\runauthor{A. P\'erez P\'erez}
\def\shorttitle{SuperB project}
\begin{document}

\title{The Super-B Factory: \\
Physics prospects and project status
}

\author{Luis~Alejandro~P\'erez~P\'erez \\
on behalf of the SuperB Collaboration}

\address{INFN - Sezione di Pisa\\
Largo Bruno Pontecorvo 3, 56127 Pisa - Italy\\
E-mail: luis.alejandro.perez@pi.infn.it}

\maketitle

\abstracts{I will briefly review of some highlights of the SuperB 
physics programme, the status of the accelerator and detector studies, and the
future plans.
}

\section{Introduction}
The LHC brought the high energy particle physics community to a new era in New Physics
(NP) searches beyond the Standard Model (SM). Direct searches of NP up to the 1 TeV scale are possible
with the ATLAS and CMS experiments, where the production of new particles and theirs decays
can be studied to constraint the NP Lagrangian. However, another complementary approach, where
high energies are substituted by high statistics, can be made by performing precision measurements
on the flavor sector of the SM. This approach is essential for two reasons: 1) New particles can manifest
themselves through small virtual effects in decays of the SM particles such as B and D mesons, and $\tau$
leptons. These contributions depend on the energy scale of the NP, hence sufficiently
precise measurements can allow to extend the NP searches beyond the 1 TeV scale. 2) These kind
of searches can be used to explore the parameter space of the NP weak complex couplings and to
determine its flavor structure.

During the last decade the SM description of the heavy flavor sector has surpassed all expectations.
The first generation of B factory experiments, BABAR and Belle, together with CDF and
D0 at TEVATRON, have provided a plethora of measurements that substantially confirm the CKM
picture of the SM, reducing strongly the parameter space of possible NP flavor mixing scenarios.
Given the high expected number of B and D mesons to be produced at the LHC, it is evident the
role of the LHCb in the NP searches program, where will be performed precise measurements of
CP-violating/CP-conserving observables from the study of rare charm and beauty decays.

The recent approval of the SuperB project open new possibilities for the flavor physics
program. The new SuperB final focusing scheme for the $e^+e^−$ collider will increase the instantaneous
luminosity by two orders of magnitude, passing from $10^{34}$ to $10^{36}~{\rm cm}^{-2}{\rm s}^{-1}$. SuperB
intends to search for indirect and some direct signs of NP at low energy, while at the same time,
enabling precision tests of the SM. The main focus of the physics programme
rests in the study of so-called {\it Golden Modes} ({\it e.g.} $B^{\pm} \rightarrow \ell^{\pm}\nu_{\ell}$, 
$B \rightarrow K^{(*)}\nu\bar{\nu}$, $B \rightarrow K^{(*)}\ell\ell$, $\tau \rightarrow 3\ell$, 
$\tau \rightarrow \mu\gamma$ and many others), these are decay channels that provide access to 
measurements of theoretically clean observables that can provide both stringent constraints on 
models of NP, and precision tests of the SM. SuperB will accumulate $75~{\rm ab}^{-1}$ of data over 
a period of five years of nominal data taking at the $\Upsilon(4S)$. In addition to operating at the 
center of mass energy of the $\Upsilon(4S)$, this experiment will also run at other energies ranging 
from charm threshold, at the $\psi(3770)$, up to the $\Upsilon(5S)$. 
The NP sensitive observables that SuperB will measure are complementary, and in many cases competitive, 
to those accessible by LHCb. Only by measuring the full set of observables at $e^+e^-$ and hadron colliders 
(LHC) will be able to optimally elucidate details of the flavor structure at high energies. SuperB will play 
a crucial role in defining the landscape of flavour physics over the next 20 years.

The physics potential~\cite{SuperB_Complementarity,SuperB_physics}, and the detector~\cite{SuperB_det} and 
accelerator~\cite{SuperB_acc} plans have been extensively documented. The accelerator parameters have been 
defined for operating in the $\psi(3770)$ to $\Upsilon(5S)$ energy range and the accelerator will reuse large 
parts of the SLAC PEP-II hardware. The project Technical Design Report (TDR) is expected to be released by the 
end of 2012. This experiment will be built at a new laboratory on the Tor Vergata campus near Rome, Italy named 
after Nicola Cabibbo. Data taking should begin five to six years after construction begins.

\section{SuperB Physics Highlights}
\subsection{$\tau$ Physics}
The intrinsic level of charged lepton flavour violation in the $\tau$ sector arising
from neutrino oscillations is expected to occur at the level of $10^{-54}$. Given
that both quark and neutral lepton number conservation is known to be violated at a 
small level, it is natural to presume that there may ultimately be non-conservation 
of charged lepton number. Indeed many NP scenarios predict large (up to $\sim 10^{-9}$) 
levels of charged lepton flavour violation (LFV). These predictions are model dependent: 
some models favour large $\mu \rightarrow e$ transitions over other possibilities, and other 
models prefer large $\tau \rightarrow \mu$ or $\tau \rightarrow e$. While the quest for a 
discovery of LFV continues, it is clear that all three sets of transitions need to be measured 
or well constrained in order to understand the underlying dynamics. SuperB will be able to 
improve existing limits from the B factories by between one and two orders of magnitude. 
Channels such as $\tau \rightarrow \ell\gamma$ will see a factor of ten improvement as these 
have irreducible SM backgrounds that one will have to contend with. Other channels such 
as $\tau \rightarrow 3\ell$, which are free of SM backgrounds, will see a factor of one 
hundred improvement. The $e^-$ beam at SuperB will be $80\%$ polarized, enabling one to separate 
contributions from SM-like LFV channels and other irreducible backgrounds as one can use 
the polarization of the final state $\tau$ lepton to suppress 
background. This works well for improving limits, or indeed searching for left handed sources 
of NP. One can verify if there is a right handed component to any underlying NP by comparing 
results with and without polarized beams.

\subsection{B Physics}
Concerning the NP search potential of SuperB, it is based on the use of indirect constraints on rare 
processes to infer the existence to the corresponding energy scale $\Lambda_{\rm NP}$ of the NP. 
The correlation between measurement of the rare decays (a branching fraction or other observable) 
and the energy scale is non trivial. 
If one considers the minimal super-symmetric model (MSSM) in the mass insertion hypothesis then for 
example the measurement of the inclusive branching fractions of $b \rightarrow s\gamma$ and 
$b \rightarrow s\ell\ell$, along with the CP asymmetry in $b \rightarrow s\gamma$ can be used to 
constrain the mass insertion parameter $(\delta^d_{23})_{LR}$. The magnitude of this parameter can be
used to infer an upper limit on $\Lambda_{NP}$ to complement the null results obtained so far from the LHC. If 
the generic MSSM is a realistic description of nature then the fact that the LHC has failed to find a 
low mass gluino implies that there is a non-trivial coupling $(\delta^d_{23})_{LR}$, and hence in 
turn SuperB should be able to observe a non-trivial deviation from the SM when studying the inclusive 
decays $b \rightarrow s\gamma$ and $b \rightarrow s\ell\ell$. The magnitude of the observed deviation 
will benefit the SLHC community as the inferred upper bound on the energy scale obtained will provide 
useful information on the integrated luminosity required to yield positive results via direct searches. 
For example if one measured $|(\delta^d_{23})_{LR}| = 0.05$, then the implied upper limit on $\Lambda_{NP}$ 
is $3.5~{\rm TeV}$, which is also compatible with known constraints on $tan\beta$ as can be seen from 
Ref.~\cite{NP_scenarios}.

There are numerous golden rare B decay channels at SuperB, including $B \rightarrow \ell\nu$, where 
$\ell = \tau, \mu, e$. In the SM this decay is known up to uncertainties related to the value of $V_{ub}$ 
and $f_B$. The rate of these processes can be modified by the existence of charged Higgs particles predicted 
in a number of extensions of the SM for example two-Higgs Doublet models (2HDM) or SUSY extensions of the SM. 
Hence the measured rate of these decays can be used to place limits on the inferred mass of any $H^+$ particle, 
and such constraints are dependent on $tan\beta$. The existing constraints from the B factories from inclusive 
$b \rightarrow s\gamma$ decays exclude masses below $295~{\rm GeV}/c^2$, and the constraint from 
$B \rightarrow \tau\nu$ excludes higher masses for large $tan\beta$ scenarios. With $75~{\rm ab}^{-1}$ of data 
SuperB will be able to exclude, or detect, a $H^+$ with a mass $1 − 3~{\rm TeV}$, for $tan\beta$ between 40
and 100. This constraint results from a combination of $B \rightarrow \tau\nu$ and $B \rightarrow \mu\nu$. 
Ref.~\cite{NP_scenarios} discusses the physics potential of several other interesting rare B decays.

Many of the CP asymmetry observables of $B_{u,d}$ decays available at SuperB are dominated by loop contributions 
and are sensitive to the same sources of NP that can affect many of the interesting rare decays discussed above. 
The golden modes to be used for measuring the angle $\beta$ of the unitarity triangle are B decays to charmonium 
($c\bar{c}$), $\eta'$ or $\psi$, and a neutral kaon. SuperB will be able to measure the CP asymmetries in these 
decays with precisions of 0.002, 0.008, and 0.021, respectively, using a data sample of $75~{\rm ab}^{-1}$. Both 
tree ($c\bar{c}K^0$) and penguin dominated decays can be affected by the presence of NP. To complement the $B_{u,d}$ 
programme at SuperB, there will be a dedicated run at the $\Upsilon(5S)$ resonance which enables the study of a 
number of $B_s$ related observables that may be affected by physics beyond the SM. These include the
semi-leptonic asymmetry and branching fraction $B_s \rightarrow \gamma\gamma$.

\subsection{Charm Physics}

Charm mixing has been established by the B factories and is parameterized by two small numbers: 
$x = \Delta m_{D}/\Gamma$ and $y = \Delta\Gamma/2\Gamma$. These are currently measured as $x = (0.65^{+0.18}_{-0.19})\%$, 
and $y = (0.74 \pm 0.12)\%$~\cite{HFAG}. The precision with which these mixing parameters can be improved upon is dominated 
by inputs from $D^0 \rightarrow K^0_Sh^+h^-$ decays ($h = \pi, K$). At large integrated luminosities
one of the limiting factors of this analysis will be the knowledge of the strong phase variation across the $K^0_Sh^+h^-$ 
Dalitz plot. This phase can be measured using data collected at the charm threshold, where $e^+e^- \rightarrow \psi(3770) \rightarrow D^0\bar{D}^0$
transitions result in pairs of quantum correlated neutral D mesons. These correlated mesons can be used to precisely determine 
the required map of the strong phase difference required for the charm mixing measurements. With this input from a data sample of 
$500~{\rm fb}^{-1}$ the mixing measurements at SuperB will still be statistics limited, and one should be able to achieve precisions 
of $0.02\%$ and $0.01\%$ on $x$ and $y$, respectively. The strong phase difference map measured at the charm threshold will also be an 
important input used for the determination of the unitarity triangle angle $\gamma$ for SuperB, Belle II and LHCb.

In the framework of the SM one expects very small CP-violation (CPV) on the charm sector, so any large measured deviation from zero 
would be a clear sign of NP. Just like the $B_{u,d}$ system, the charm sector has a unitarity triangle that needs to be tested. The 
physics potential of SuperB in this area has recently been outlined in Ref.~\cite{BIM}. In the months following the Lomonosov 
conference an intriguing hint of CPV in charm decays was produced by the LHCb experiment~\cite{LHCb_Acp_D}. This relates to a 
difference in direct CP asymmetry parameters measured in $D \rightarrow KK$ and $D \rightarrow \pi\pi$. If this is a real effect 
one will have to perform the measurements outlined in~\cite{BIM} in order to understand the underlying physics.

\subsection{Interplay Between Measurements}

The power of SuperB comes from its ability to study a diverse set of modes that are sensitive to different types of NP. Through 
the pattern of deviations from the SM expectations for the sensitive observables one will be able to identify viable NP scenarios and reject those 
that are not compatible with the data. This goes beyond the motivation of simply discovering some sign of NP and is a step toward 
developing a detailed understanding of NP. If no significant deviations are uncovered then this in turn can be used to constrain 
parameter space and reject models that are no longer viable. Given that many of the observables that SuperB will measure are not 
accessible directly at the LHC, these results will complement the direct and indirect searches being performed at LHC.
Detailed discussions on the interplay problem can be found in Refs.~\cite{SuperB_Complementarity,NP_scenarios}.

\section{SuperB Accelerator Highlights}

The SuperB collider exploits a novel collision scheme~\cite{Pantaleo}, based on very small beam 
dimensions and betatron function at the interaction point, on large crossing and Piwinsky angle
and on the “crab waist” scheme. This approach allows to reach the required luminosity of $10^{36}~{\rm cm}^{-2}{\rm s}^{-1}$
and at the same time overcome the difficulties of early super $e^+e^-$ collider designs, most notably
very high beam currents and very short bunch lengths. The wall-plug power and the beam-related
background rates in the detector are therefore kept within affordable levels~\cite{SuperB_acc}.
The crab waist transformation consists in moving the waist of each beam onto the axis of the
other beam with a pair of sextupole up- and down-stream the interaction point. In this way all particles from
both beams collide in the minimum $\beta^*_y$ region, with a net luminosity gain. Moreover (and most
significantly) the $x/y$ betatron resonances are naturally suppressed. The principle of the innovative
interaction region (IR) design sketched above has been experimentally demonstrated at the Frascati DA$\Phi$NE collider~\cite{DAFNE}.
It is very importantly that, this test also validated the simulations used to calculate the IR optics.
The SuperB design is based on recycling as much as possible the existing PEP-II hardware, with
a significant reduction of costs. The optimal beam energy choice for the accelerator design, 
is $4.18~{\rm GeV}$ electron beam (polarization of $80\%$) and $6.71~{\rm GeV}$ positron beam.
The low currents, ultra-small emittance approach has been adopted recently also by the KEKB
accelerator team, which defined a new set of parameters very similar to that of the Italian SuperB.

\section{SuperB Detector Highlights}

Most of the general requirements for the SuperB detector are common to those of the present B
factories, including large solid angle coverage, good particle identification (PID) capabilities over
a wide momentum range ($\pi/K$ separation to over $4~{\rm GeV/c}$), measurement of the relative decay
times of the B mesons, good resolution of the charged track momentum and of the photon
energy, particularly in the sub-${\rm GeV}$ part of the spectrum, relevant at the $\Upsilon(4S)$ 
environment. The SuperB detector concept is therefore based on the BABAR detector, with the modifications 
required to operate at a much higher luminosity (and luminosity-scaling background rates), and with a reduced
center-of-mass boost~\cite{SuperB_det}.

The BABAR detector is composed by a tracking system – a five layer double-sided silicon strip vertex tracker 
(SVT) and a 40 layer drift chamber (DCH) immersed in a $1.5~{\rm T}$ magnetic field – a Cherenkov detector with 
fused silica bar radiators (DIRC), a homogeneous electromagnetic calorimeter made of CsI(Tl) crystals (EMC), 
and a detector for muon identification and $K^0_L$ detection (IFR) realized instrumenting the iron flux return 
with resistive plate chambers and limited streamer tubes. SuperB is designed to reuse a number of BABAR components: 
the DIRC quartz bars, the CsI(Tl) crystals of the barrel EMC, the flux-return steel, the superconducting coil.

The center-of-mass boost at SuperB is smaller than in BABAR ($\beta\gamma = 0.24$ vs. $0.56$). While this effectively 
improves the angular coverage of the detector, it also reduces the $\Delta z$ separation of the decay vertices. 
The $\Delta t$ sensitivity in time-dependent measurements is maintained by improving the vertex resolution: the 
SuperB vertex detector replicates the five-layer BABAR SVT, but exploits the reduced dimensions of the beam pipe 
made possible by the ultra-low emittance SuperB beams to add a very thin and precise measurement layer at a radius 
of only $1.6~{\rm cm}$. The baseline technology for this “Layer0” uses short double-sided silicon strip detectors 
(“striplets”), while other options (pixel silicon sensors) are being considered as possible upgrades. The SuperB DCH concept is derived from 
the BABAR one, with several improvements. The hadron PID system will use the radiator quartz bars of the BABAR DIRC, 
read-out by fast multi-anode PMTs, and with the imaging region considerably reduced in size to improve performance and 
reduce the impact of backgrounds. The forward EMC will feature cerium-doped LYSO crystals, which have a much shorter 
scintillation time constant, a smaller Molière radius and better radiation hardness than the current CsI(Tl) crystals, 
for reduced sensitivity to beam backgrounds and better position resolution. The thickness of the flux-return iron will 
be increased with an additional absorber to bring to about 7 the number of interaction lengths for muons, while the gas 
detectors will be replaced by extruded plastic scintillator bars to cope with the expected background rates. Finally,
the Collaboration is considering to improve the detector hermeticity by inserting a “veto-quality” lead-scintillator 
EMC calorimeter in the backward direction, and to add a particle identification device in front of the forward calorimeter. 

\section{Summary}

The physics programme at SuperB is varied, and the unique features of the facility: polarized electron beams and 
a dedicated charm threshold run add to its strengths via versatility. The charm threshold run in particular, in addition to 
facilitating a number of NP searches, will provide several measurements required to control systematic uncertainties for 
measurements of charm mixing and the unitarity triangle angle $\gamma$. Results from SuperB will surely play a role in 
elucidating any NP discovered at the LHC and indirectly probe to higher energy than the LHC will be able to directly access.

\end{document}